\documentclass[aps,prl,superscriptaddress,reprint]{revtex4-2}
\usepackage{amsfonts}
\usepackage{mathrsfs}
\usepackage{xcolor}
\usepackage{graphicx}
\usepackage{mathtools}
\usepackage{etoolbox}
\usepackage{setspace}

\usepackage[stretch=10,shrink=30,step=1]{microtype}
\usepackage[english]{babel}
\renewcommand{\eqref}[1]{Eq.~(\ref{#1})}

\newcommand{\Phieff}{\mathcal{E}}

\begin{document}

\title{
Phase-Space Quantum Geometry Beyond Adiabatic Electron Dynamics
}

\author{Luca Maranzana}
\email[]{Luca.Maranzana@iit.it}
\affiliation{Italian Institute of Technology, Via Morego 30, Genoa, Italy}
\affiliation{Department of Physics, University of Genoa, Via Dodecaneso 33, Genoa, Italy}
\affiliation{Quantum Materials Theory, 16162, Genoa, Italy}

\author{Koki Shinada}
\affiliation{RIKEN Center for Emergent Matter Science (CEMS), Wako, Saitama 351-0198, Japan}

\author{Ying-Ming Xie}
\altaffiliation[Present address: ]{Institute of Condensed Matter Physics, School of Physics and Astronomy, Shanghai Jiao Tong University, Shanghai, China}
\affiliation{RIKEN Center for Emergent Matter Science (CEMS), Wako, Saitama 351-0198, Japan}

\author{Sergey Artyukhin}
\affiliation{Quantum Materials Theory, 16162, Genoa, Italy}

\author{Naoto Nagaosa}
\email[]{nagaosa@riken.jp}
\affiliation{RIKEN Center for Emergent Matter Science (CEMS), Wako, Saitama 351-0198, Japan}
\affiliation{Fundamental Quantum Science Program (FQSP), TRIP Headquarters, RIKEN, Wako 351-0198, Japan \looseness=-1}

\begin{abstract}
    The geometry of electronic quantum states plays an important role in the equilibrium and transport properties of solids. While the Berry curvature is known to influence electron motion, recent work has shown that the quantum metric also affects the motion of electron wave packets beyond the adiabatic approximation. To connect this nonadiabatic dynamics to many-electron observables, we derive an equivalent semiclassical formulation, valid up to second order in $\hbar$. The resulting phase-space measure and kinetic equation incorporate the quantum metric over the full phase space, including its mixed real-momentum components. We show that, in spatially inhomogeneous systems, the full phase-space quantum metric contributes to electric polarization in insulators and generates an intrinsic linear Hall response in metals. As a concrete example, we study Dirac electrons subject to a magnetic texture and a background potential that vary periodically in space. In this model, the mixed components of the phase-space metric produce a Hall contribution controlled by the relative phase between the two modulations. This phase-sensitive response can remain finite even when the conventional anomalous Hall conductivity vanishes. More broadly, our formulation enables the systematic study of equilibrium and transport responses in systems whose quantum geometry involves both position and momentum.

\end{abstract}

\maketitle

Semiclassical wave-packet dynamics provides a bridge between the microscopic quantum mechanics of electrons and the macroscopic properties of solids. In this approach, an electron is represented as a localized wave packet built from Bloch states, whose center evolves in real and momentum space according to semiclassical equations of motion \cite{Chang96, Sundaram99, Xiao10}. Importantly, this dynamics depends not only on the band energy but also on the quantum geometry of Bloch states. Momentum-space quantum geometry characterizes how Bloch states change with crystal momentum. In inhomogeneous crystals, the local Bloch states can also vary with position. The quantum geometry then extends over the full phase space and can include real-space and mixed components.

Berry curvature is the best-known manifestation of quantum geometry \cite{Berry84, Zak89, Karplus54, Chang96, Sundaram99, Xiao05, Xiao10, Nagaosa10, Sodemann15, Ahmad26, Xiao09}. It generates an anomalous velocity, a transverse contribution to electron motion that underlies the intrinsic anomalous Hall effect \cite{Karplus54, Nagaosa10}. It also modifies the phase-space measure, which determines the number of available quantum states in a given phase-space volume, with consequences for both transport and thermodynamic properties \cite{Xiao05, Xiao10}. Complementary to the Berry curvature, the quantum metric measures the distance between nearby quantum states. Beyond this geometric interpretation, it plays a key role in nonlinear transport, electric multipolar responses, and orbital magnetism \cite{Yu25, Gao25, Gao14, Gao15, Gao19, Gao23, Lapa19, Daido20, Watanabe21, Zhao21, Xiao21, Ahn20, Ahn22, Mehraeen25, Nagaosa22, Kozii21, Hetenyi23, Shinada23, Shinada25, Xie25, Sala25, Sala26, Komissarov24, Kaplan24, Ulrich25, Mitscherling25, Qiang25}. The quantum metric also enters nonadiabatic wave-packet dynamics, in which the electronic state cannot adjust instantaneously as the wave packet moves through phase space \cite{Smith22, Shinada26, Onishi25, Yoshida25, Ren25A, Ren25B}.

Connecting the nonadiabatic dynamics of a single wave packet to many-electron observables remains challenging. In nonlinear response theory, quantum-metric effects are typically encoded in field-induced corrections to the Berry connection and wave-packet energy \cite{Gao14, Gao15}. In nonadiabatic wave-packet dynamics, however, the quantum metric enters in a form less suited to kinetic theory: the equations of motion contain second time derivatives of phase-space coordinates \cite{Onishi25, Yoshida25, Ren25A, Ren25B}. This structure makes it challenging to derive the phase-space measure and the kinetic equation needed to calculate many-electron observables. Existing derivations of nonadiabatic corrections to the phase-space measure are restricted to pure momentum-space geometry and, hence, omit real-space and mixed metric components \cite{Mameda25}. The mixed components, in particular, can produce response contributions that cannot be captured by treating real- and momentum-space geometries separately.

\begin{figure*}[t]
\includegraphics[width=0.99\linewidth]{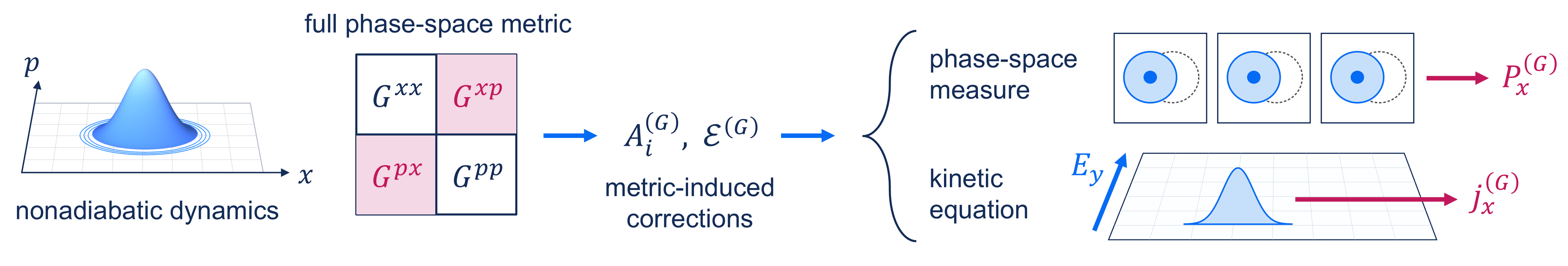}
    \caption{\label{fig:Fig1} In the semiclassical description, an electron is represented by a wave packet whose center evolves in phase space. Its nonadiabatic dynamics involves the band-renormalized quantum metric $G_{ij}$, which extends over full phase space and contains real-space, momentum-space, and mixed components, with the latter highlighted in magenta. The metric gives rise to corrections $A_i^{\smash{(G)}}\!$ and $\Phieff^{\smash{(G)}}\!$ to the Berry connection and wave-packet energy, respectively. These corrections enter the phase-space measure and kinetic equation. In insulators, the correction to the measure produces a polarization contribution $P_x^{\smash{(G)}}\!$. In metals, an applied field $E_y$ drives an intrinsic linear Hall current $j_x^{\smash{(G)}}\!$. In general, both responses can receive contributions from the momentum-space and mixed components of $G_{ij}$.}
\end{figure*}

To enable a consistent kinetic description in full phase space, we derive an equivalent formulation of nonadiabatic wave-packet dynamics. The resulting equations of motion involve only first time derivatives and reproduce the original dynamics up to second order in $\hbar$. In this framework, quantum-metric effects appear as corrections to the Berry connection and wave-packet energy. This structure allows a systematic derivation of the corresponding phase-space measure and kinetic equation. The main elements of this construction are summarized in Fig.~\ref{fig:Fig1}.

We next consider the physical consequences of retaining the full phase-space geometry. For a completely filled band, the nonadiabatic contribution to the phase-space measure generates a polarization involving both momentum-space and mixed components of the band-renormalized quantum metric. This mechanism is distinct from that identified in Ref.~\cite{Zhao21}, where polarization arises from spatial variations of the bare momentum-space metric. Turning to transport in spatially inhomogeneous metals, we identify an intrinsic linear Hall contribution arising from the full phase-space metric. As a concrete example, we consider Dirac electrons coupled to a magnetic texture and a background potential that vary periodically in space. In this model, the mixed components of the phase-space metric produce an intrinsic Hall contribution controlled by the relative phase between the two modulations. This contribution can remain finite even when the conventional anomalous Hall effect vanishes.

\section{From nonadiabatic wave-packet dynamics to kinetic theory}
We begin with the Lagrangian describing the nonadiabatic dynamics of a wave packet in a nondegenerate Bloch band \cite{Ren25A, Ren25B, Yoshida25, Mameda25}
\begin{equation}\label{eq:L}
	L = \frac{\hbar^2}{2} G_{ij}\dot{\xi}^i \dot{\xi}^j + \dot{\xi}^i \left(\hbar A_i + \frac{1}{2}J_{ij}\xi^j\right) - \Phieff,
\end{equation}
where $\boldsymbol{\xi} = (\boldsymbol{x},\boldsymbol{p})$ is the phase-space coordinate of the wave packet center, the indices run through the full phase space, and the Einstein summation convention is implied. $G_{ij}(\boldsymbol{\xi})$ is the band-renormalized quantum metric, $A_i(\boldsymbol{\xi})$ the Berry connection, $\Phieff(t,\boldsymbol{\xi})$ the energy of the wave packet, and $J_{ij}$ the canonical symplectic matrix:
\begin{equation}\label{eq:J}
	J = \begin{pmatrix} 0 & I_d \,\\ -I_d & 0 \,\end{pmatrix}\!,
\end{equation}
where $I_d$ is the $d \times d$ identity matrix acting on either real space or momentum space, each of dimension $d$. We allow for a weak time-dependent perturbation of the wave-packet energy, $\Phieff(t,\boldsymbol{\xi}) = \Phieff_0(\boldsymbol{\xi}) + \delta\Phieff(t,\boldsymbol{\xi})$, such as that produced by an oscillating electric field. By contrast, $G_{ij}(\boldsymbol{\xi})$ and $A_i(\boldsymbol{\xi})$ are defined from the unperturbed Bloch states (i.e., in the absence of the drive) and thus remain time-independent.

It is important to distinguish the nonadiabatic effects considered here from the field-induced corrections familiar from nonlinear response theory. An applied field modifies the Hamiltonian eigenstates and thus alters their quantum geometry \cite{Gao14}. This effect cannot be captured by the wave-packet energy alone, but requires including the gradient of the underlying Hamiltonian. Such gradient corrections are well established in the literature and have been generalized beyond the electric-field case \cite{Zhao21, Xiao21, Ren25B}. Here, instead, we focus on the nonadiabatic corrections associated with the quantum-metric term in the Lagrangian, \eqref{eq:L}.

\subsection{Effective Lagrangian and equations of motion}
The quantum-metric term in the Lagrangian makes the equations of motion second order in time. A key step in our construction is to recast the dynamics in an equivalent first-order form, valid up to $\mathcal{O}(\hbar^2)$. The corresponding effective Lagrangian is (see End Matter for the full derivation):
\begin{equation}\label{eq:Leff}
	L_{\mathrm{eff}} = \dot{\xi}^i \left(\hbar A_i^\mathrm{eff} + \frac{1}{2}J_{ij}\xi^j\right) - \Phieff^\mathrm{eff} + \mathcal{O}\!\left(\hbar^3\right)\!.
\end{equation}
The effective Lagrangian retains the standard semiclassical form \cite{Chang96, Sundaram99, Xiao10}, with the nonadiabatic quantum-metric effects encoded in the effective Berry connection $A_i^\mathrm{eff} = A_i + \hbar A_{i}^{\smash{(G)}}\!$ and wave-packet energy $\Phieff^\mathrm{eff} = \Phieff + \hbar^2 \Phieff^{\smash{(G)}}$:
\begin{equation}\label{eq:Corr}
	A_{i}^{(G)} = -G_{ij} J^{jk} \partial_k \Phieff, \quad \Phieff^{(G)} = \frac{1}{2} G_{ij} J^{ia}J^{jb} \partial_a \Phieff \partial_b \Phieff,
\end{equation}
where $J^{ik}$ is the inverse symplectic matrix: $J^{ik}J_{kj} = \delta^i_{\,j}$. These contributions resemble the field-induced corrections known from nonlinear response \cite{Gao14, Gao15}, but have a distinct physical origin. They originate from nonadiabatic effects in the wave-packet dynamics, rather than from gradients of the underlying single-particle Hamiltonian. In the special case of pure momentum-space geometry and DC driving, analogous corrections were derived in Ref.~\cite{Mameda25} within the Dirac formalism for constrained systems \cite{Dirac50, Dirac64}.

Varying the effective Lagrangian gives the compact first-order equations of motion
\begin{equation}\label{eq:EqEff}
	\dot{\xi}^i = \left(\hbar \Omega^\mathrm{eff} - J\right)^{\!ik} \left(\partial_k \Phieff^\mathrm{eff} - \hbar^2 G_{ka} J^{aj} \partial_t \partial_j \Phieff \right) + \mathcal{O}\!\left(\hbar^3\right)\!,
\end{equation}
where $\Omega_{ik}^\mathrm{eff} = \partial_{i}A_{k}^\mathrm{eff} - \partial_{k}A_{i}^\mathrm{eff}\!$ is the effective Berry curvature and $\big(\hbar \Omega^\mathrm{eff} - J\big)^{\smash{\!ik}} \big(\hbar \Omega_{kj}^\mathrm{eff} - J_{kj}\big) = \delta^i_{\,j}$. In the presence of an external electric field $\boldsymbol{E}(t)$ the wave-packet energy takes the form $\Phieff(t,\boldsymbol{\xi}) = \Phieff_0(\boldsymbol{\xi}) -e \Phi(t,\boldsymbol{x})$, where $\boldsymbol{E}(t) = -\nabla_{\!\boldsymbol{x}} \Phi$. Hence, the electric field generates the usual conservative force together with a nonadiabatic correction proportional to $\partial_t \boldsymbol{E}$, which originates from the $\partial_t \partial_j \Phieff$ term in \eqref{eq:EqEff}.

\subsection{Phase-space measure and kinetic equation}
The phase-space measure is determined by the symplectic structure associated with the effective Lagrangian. Up to second order in $\hbar$, the corresponding density of states is
\begin{equation}\label{eq:EqDhbar2}
	D = \frac{1}{(2\pi\hbar)^{d}} \!\left(1 - \frac{\hbar}{2}\Omega_{ij}^\mathrm{eff}J^{ji} + \hbar^2 \mathcal{I}_{\Omega}^{(2)} + \mathcal{O}\!\left(\hbar^{3}\right)\! \right)\!,
\end{equation}
where $\mathcal{I}_{\Omega}^{(2)}$ denotes the quadratic Berry-curvature contribution, corresponding to the second Chern form \cite{Zhao21} (see End Matter for its expression). The terms involving only the Berry curvature reproduce the standard phase-space measure \cite{Xiao05, Zhao21}, while the quantum metric enters through $\Omega_{ij}^\mathrm{eff} = \Omega_{ij} + \hbar\!\:\Omega_{ij}^{\smash{(G)}}\!$, with
\begin{eqnarray}\label{eq:EqOmG}
	\Omega_{ij}^{(G)}\! &=& \left(\partial_j G_{ik} - \partial_i G_{jk}\right) J^{ka} \partial_a\Phieff \nonumber\\
    & & + J^{ka} \left(G_{ik} \partial_j\partial_a\Phieff - G_{jk} \partial_i\partial_a\Phieff\right)\!.
\end{eqnarray}
The two contributions to $\Omega_{ij}^{\smash{(G)}}\!$ originate, respectively, from phase-space derivatives of the quantum metric and from its coupling to second derivatives of the wave-packet energy. The density of states $D$ provides the semiclassical number of quantum states per unit of phase-space volume and is inversely proportional to the minimal uncertainty volume \cite{Xiao05}. Therefore, the quantum-metric contribution in \eqref{eq:EqOmG} is essential to construct the phase-space measure consistently through $\mathcal{O}(\hbar^2)$ and to evaluate thermodynamic and transport observables at the same order.

With the modified density of states $D$, the phase-space probability density becomes $\rho(t,\boldsymbol{\xi}) = D(t,\boldsymbol{\xi}) f(t,\boldsymbol{\xi})$, where the distribution function $f$ satisfies the kinetic equation
\begin{equation}\label{eq:EqB}
    \partial_t f + \dot{\xi}^i \partial_i f = I_\mathrm{coll} \approx -\frac{f-f_0}{\tau}.
\end{equation}
Here $f_0$ is the equilibrium distribution function, $I_\mathrm{coll}$ the collision integral within the relaxation-time approximation, and $\tau$ the relaxation time. Since the equilibrium distribution function includes the energy correction, \eqref{eq:Corr}, we expand it as $f_0(\Phieff_0 + \hbar^2\Phieff^{(G)}) = f_0(\Phieff_0) + \hbar^2\Phieff^{(G)}\! f'_0(\Phieff_0)$ \cite{Qiang25}, where the prime denotes differentiation with respect to the unperturbed band energy. Together, \eqref{eq:EqEff}, \eqref{eq:EqDhbar2}, and \eqref{eq:EqB} connect the nonadiabatic dynamics of an individual wave packet to a kinetic description of a many-electron system.

As a consistency check, for pure momentum-space geometry and a static electric field, our framework reproduces the results of Ref.~\cite{Mameda25} (see End Matter). Remarkably, the resulting current matches the expression obtained from the field-induced corrections \cite{Gao14}. This correspondence arises because, in this specific case, the nonadiabatic corrections in \eqref{eq:Corr} coincide with their field-induced counterparts. Beyond this limit, the full phase-space formulation reveals additional contributions arising from spatial inhomogeneity, including those involving the mixed real-momentum components of the quantum metric.

\section{Equilibrium responses from full phase-space geometry}
We now examine the equilibrium responses arising from the full phase-space quantum metric. At zero electric field, the energy density $\rho_\varepsilon = \int{\!d^d p}\, D \Phieff^\mathrm{eff}\! f_0(\Phieff^\mathrm{eff})$ takes the form, up to the second order in $\hbar$,
\begingroup\setlength{\arraycolsep}{0pt}\begin{eqnarray}
    \rho_\varepsilon = \!\int\!\frac{d^d p}{(2\pi\hbar)^{d}} & & \bigg\{ \Phieff_0 f_0\!\left[1-\frac{\hbar^2}{2} \partial_i\!\left(J^{ia}G_{ab}J^{bk}\partial_k \Phieff_0\right)\right] \nonumber\\
    & &- \frac{\hbar^2}{2} \partial_i\!\left(\Phieff_0 f_0J^{ia}G_{ab}J^{bk}\partial_k \Phieff_0\right)\!\bigg\} +  \rho_\varepsilon^{(\Omega)}\!, \quad \label{eq:EqE}
\end{eqnarray}\endgroup
where $f_0 \equiv f_0(\Phieff_0)$, and $\rho_\varepsilon^{\smash{(\Omega)}}$ is the Berry-curvature contribution, given explicitly in the End Matter. The quantum metric introduces corrections to $\rho_\varepsilon$ proportional to derivatives of the unperturbed band energy $\Phieff_0$, namely the group velocity $v_i = \partial_{i}^{p} \Phieff_0$ and the real-space energy gradient $\partial_{i}^{x} \Phieff_0$. Hereafter, indices run over either real or momentum space, with superscripts specifying the corresponding space.

\subsection{Metric-induced polarization}
The electric-field contribution to the energy density can be decomposed as $\rho_{\varepsilon,\!\:\mathrm{tot}}^{\smash{(E)}} = \rho_{\varepsilon}^{\smash{(E)}} - e n\Phi$. Here, $- e n\Phi$ is the standard coupling of the local charge density to the scalar potential, and $n = \int{\!d^{d} p}\, D f_0(\Phieff^\mathrm{eff})$ is the particle-number density. Since $n$ includes the geometric corrections to the phase-space measure, the local charge density acquires a spatial-divergence contribution. In an insulator, Ref.~\cite{Zhao21} has shown that this contribution can be identified with a polarization charge. In particular, any field-independent term $a_i^p$ in the momentum-space Berry connection induces a polarization $P_i^{\smash{(a)}} \propto -e\int{\!d^d p}\,a_i^p$, in accordance with the modern theory of polarization \cite{Vanderbilt93, Resta94}. Consequently, the nonadiabatic correction $A_i^{p\!\:\smash{(G)}}\!$, evaluated at zero electric field, gives rise to the polarization contribution
\begin{equation}\label{eq:PG}
	P_{i}^{(G)} = e\hbar^2\!\int\!\frac{d^d p}{(2\pi\hbar)^{d}} \!\left(G_{ij}^{pp} \partial_j^x \Phieff_0 - G_{ij}^{px} v_j\right)\!.
\end{equation}
Since the bulk polarization is defined here for an insulator, we set $f_0 = 1$ in \eqref{eq:PG}. Elsewhere, we retain a generic $f_0$ to describe both metals and insulators. The two terms in \eqref{eq:PG} involve different components of the quantum metric. The first couples the momentum-space metric $G_{ij}^{pp}$ to the real-space gradient of the unperturbed band energy, while the second couples the mixed real-momentum metric $G_{ij}^{px}$ to the group velocity. Since both $\partial_j^x\Phieff_0$ and $G_{ij}^{px}$ vanish in the homogeneous limit, the metric-induced polarization $P_{i}^{\smash{(G)}}$ requires spatial inhomogeneity.

\subsection{Quantum-geometric quadrupole moment density}
We now focus on $\rho_{\varepsilon}^{\smash{(E)}}$, which contains the electric-field-dependent corrections to the phase-space density of states and wave-packet energy (see End Matter for their explicit expressions). In addition to the nonadiabatic
corrections, the finite spatial extent of the wave packet produces the quadrupolar energy contribution, $\Phieff^{(q)} = (e\hbar^2/2)\!\:g_{ij}^{pp}\partial_i^x E_j$ \cite{Zhao21}. Here, $g_{ij}^{pp}$ denotes the bare momentum-space quantum metric, which is distinct from the band-renormalized metric $G_{ij}$ entering the nonadiabatic corrections derived above.
Combining all these contributions and expanding the equilibrium distribution as discussed above, we obtain
\begingroup\setlength{\arraycolsep}{0pt}\begin{eqnarray}
    \rho_\varepsilon^{(E)} = & & \; \frac{e^2\hbar^2}{2} \!\int\!\frac{d^d p}{(2\pi\hbar)^{d}}\, G_{ij}^{pp} E_i E_j \bigg(1 + \Phieff_0 \frac{\partial}{\partial\Phieff_0}\bigg) f_0 \nonumber\\
    & & + \left(q_{ij} + Q_{ij}\right)\partial_i^x E_j + E_j \partial_i^xQ_{ij}, \label{eq:EqEE}
\end{eqnarray}\endgroup
with two terms in the quadrupole-moment density
\begin{eqnarray}
    q_{ij} &=& \frac{e\hbar^2}{2}\!\int\!\frac{d^d p}{(2\pi\hbar)^{d}}\, g_{ij}^{pp} \bigg(1 + \Phieff_0 \frac{\partial}{\partial\Phieff_0}\bigg) f_0, \label{eq:Qg}\\
    Q_{ij} &=& e\hbar^2 \!\int\!\frac{d^d p}{(2\pi\hbar)^{d}}\, G_{ij}^{pp} \Phieff_0 f_0. \label{eq:Q}
\end{eqnarray}
In addition to the terms quadratic in the electric field and proportional to its gradient \cite{Mameda25, Zhao21}, $\rho_\varepsilon^{\smash{(E)}}\!$ contains a term linear in $E_j$. Although this term combines with $Q_{ij} \partial_i^x E_j$ into a divergence, it gives a finite local contribution to the energy density. In an insulator, it can be interpreted as an additional contribution to the metric-induced polarization in \eqref{eq:PG}. Notably, all contributions involving the mixed phase-space metric $G_{ij}^{px}$ cancel, so that $\rho_\varepsilon^{\smash{(E)}}\!$ involves only the momentum-space metrics $g_{ij}^{pp}$ and $G_{ij}^{pp}\!$.

At zero temperature, the grand-potential density takes the form $\rho_g = \rho_\varepsilon - \mu\!\:n$, where $\mu$ is the chemical potential. The electric-field contribution $\rho_g^{\smash{(E)}}$ is obtained from Eqs. (\ref{eq:EqEE}--\ref{eq:Q}) by the replacement $\Phieff_0 \to \Phieff_0 - \mu$. As a result, the quadrupole-moment density becomes
\begin{equation}\label{eq:Qtot}
	\mathcal{Q}_{ij}^\mathrm{tot} =e\hbar^2\!\int\!\frac{d^d p}{(2\pi\hbar)^{d}} \!\left( \frac{1}{2}\, g_{ij}^{pp} + \left(\Phieff_0 - \mu\right) G_{ij}^{pp}\right)\!f_0 ,
\end{equation}
where we use $\left(\Phieff_0 - \mu\right) f'_0 = -\left(\Phieff_0 - \mu\right) \delta\!\left(\Phieff_0 - \mu\right)=0$. We note that \eqref{eq:Qtot} includes both the quantum-geometric contributions 
to the quadrupole-moment density discussed in Ref.~\cite{Daido20},
which adopts the full quantum approach under the assumption of spatially uniform geometry. In contrast, Ref.~\cite{Zhao21} includes only the contribution in \eqref{eq:Qg}. Thus, our analysis extends the conventional semiclassical wave-packet description of Ref.~\cite{Zhao21} by taking into account the nonadiabatic effects from the band-renormalized quantum metric. Furthermore, it provides a consistent formalization for non-uniform quantum geometry.
The quantum-kinetic theory offers a complementary approach to studying polarization in spatially inhomogeneous systems \cite{Suzuki}.



\section{Metric-induced linear Hall response}
We now turn from equilibrium properties to linear DC transport. At first order in the electric field, the current $j_i = -e\int{\!d^d p}\, D \dot{x}_i f$ contains an intrinsic contribution from the quantum metric,
\begin{equation}\label{eq:jG}
	j_{i}^{(G)} = -e^2 \hbar^2 E_k \!\int\!\frac{d^d p}{(2\pi\hbar)^{d}}\, \Omega_{ik}^{pp\!\:\smash{(G)}}\!\big|_{\Phieff_0} f_0 + j_{i}^{(M)}\!,
\end{equation}
where the effective Berry curvature in momentum space $\Omega_{ik}^{pp\!\:\smash{(G)}}\!$, \eqref{eq:EqOmG}, is evaluated with the unperturbed band energy $\Phieff_0$. The second term, $j_{i}^{\smash{(M)}}\!$, denotes a magnetization current. In three dimensions, $\boldsymbol{j}^{\smash{(M)}}\! = \nabla_{\!\boldsymbol{x}} \times \boldsymbol{M}$, where
\begin{equation}\label{eq:jM}
	M_i = e^2 \hbar^2 E_k \!\int\!\frac{d^d p}{(2\pi\hbar)^{d}}\, \varepsilon_{ijl} G_{jk}^{pp}  v_l f_0,
\end{equation}
is the corresponding local orbital magnetization. The first term in \eqref{eq:jG} has the same structure as the anomalous Hall current, but with the metric-induced effective Berry curvature $\Omega_{ik}^{pp\!\:\smash{(G)}}\!$. This curvature contains both the mixed metric, $G_{ij}^{px}$, and the momentum-space metric, $G_{ij}^{pp}$. The latter contribution is nevertheless a full phase-space effect, because it is tied to the real-space dependence of $\Phieff_0$. For purely momentum-space geometry and a spatially uniform $\Phieff_0$, $\Omega_{ik}^{pp\!\:\smash{(G)}}\!$ vanishes.  Under these conditions, $j_{i}^{\smash{(G)}}\!$ reduces to a pure magnetization current of the form discussed in Refs.~\cite{Lapa19, Mameda25}. The associated local orbital magnetization coincides with that obtained by applying the formalism of Ref.~\cite{Xiao21} to the nonadiabatic corrections in \eqref{eq:Corr}.

\begin{figure*}[t]
\includegraphics[width=0.99\linewidth]{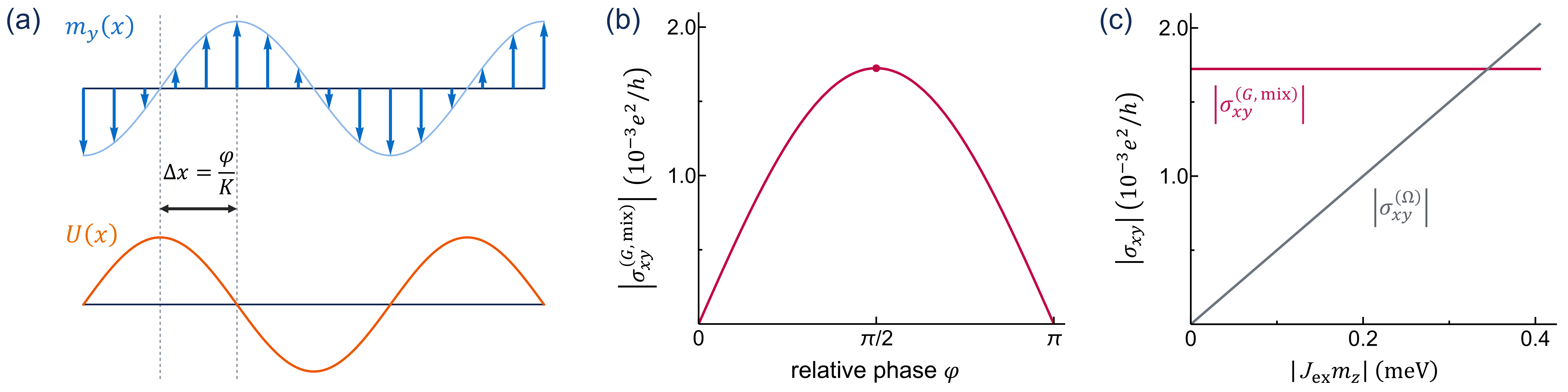}
    \caption{\label{fig:Fig2} Mixed-metric Hall response of two-dimensional Dirac electrons subject to a magnetic texture and a scalar potential. (a) The in-plane magnetization component $m_y(x) = m_y\cos(Kx)$ and the scalar potential $U(x) = u\cos(Kx+\varphi)$ have a relative phase $\varphi$, corresponding to a spatial shift $\Delta x=\varphi/K$ between the two modulations. (b) In the metallic regime, the magnitude of the mixed-metric Hall conductivity $\lvert\sigma_{xy}^{\smash{(G,\!\:\mathrm{mix})}}\rvert$ is proportional to $\lvert\sin\varphi\rvert$. Hence, it vanishes when the two modulations are in phase or antiphase and reaches its maximum for a quarter-period shift. (c) At the order retained here, $\sigma_{xy}^{\smash{(G,\!\:\mathrm{mix})}}\!$ is independent of the Dirac mass $J_\mathrm{ex} m_z$ generated by the uniform out-of-plane magnetization. For $\varphi=\pi/2$, it remains finite at $m_z = 0$, whereas the conventional intrinsic anomalous Hall conductivity $\sigma_{xy}^{\smash{(\Omega)}}\!$ vanishes linearly as $m_z \to 0$.
    }
\end{figure*}

For a periodically modulated system, the spatial average of the magnetization current vanishes. The metric-induced bulk conductivity is therefore
\begin{equation}\label{eq:sigmaG}
	\sigma_{ij}^{(G)} = -\frac{e^2 \hbar^2}{V} \!\int\!\frac{d^{2d} \xi}{(2\pi\hbar)^{d}}\, \Omega_{ij}^{pp\!\:\smash{(G)}}\!\big|_{\Phieff_0} f_0,
\end{equation}
where $V$ is the volume of the system. The conductivity in \eqref{eq:sigmaG} is antisymmetric under $i \!\leftrightarrow\! j$ and hence describes a Hall-type response. This contrasts with the contribution of the ordinary mixed Berry curvature $\Omega_{ij}^{xp}$, which does not generate a $\tau^{\smash{0}}$ response and, at $\tau^{\smash{1}}$, yields only a symmetric (i.e., non-Hall) conductivity \cite{Yokoyama21}. We also note that $\sigma_{ij}^{\smash{(G)}}$ is odd under time reversal, as required for a Hall response. This follows from \eqref{eq:EqOmG}: the indices contracted through $J^{ka}$ necessarily belong to conjugate real- and momentum-space directions, so that $\Omega_{ij}^{pp\!\:\smash{(G)}}\!$ transforms
with the same time-reversal parity as the ordinary Berry curvature.


The analysis above isolates the quantum-metric correction associated with nonadiabatic dynamics, which is the focus of the present study. Gradient-induced corrections to the Berry connection and wave-packet energy have been derived in previous works \cite{Gao14, Zhao21, Xiao21, Ren25B} and are therefore not rederived here. In Ref.~\cite{Ren25B}, the Lagrangian \eqref{eq:L} is obtained within a formalism that also includes a gradient perturbation $H_1$. We denote the corrections arising from its interband matrix elements by $A_{i}^{\smash{(H_1)}}\!$ and $\Phieff^{\smash{(H_1)}}\!$, respectively. Their explicit expressions are collected in the End Matter for completeness. At the order retained here, the $H_1$-induced and metric-induced corrections enter the effective Berry connection and wave-packet energy as separate additive terms. Hence, the $H_1$-induced Hall contribution is obtained from \eqref{eq:sigmaG} by replacing the metric-induced curvature correction with its $H_1$ counterpart $\Omega_{ij}^{pp\!\:\smash{(H_1)}}\!$. In the following section, we consider a model where this contribution exactly cancels the part of $\sigma_{ij}^{\smash{(G)}}\!$ arising from $G_{ij}^{pp}$, leaving a Hall response controlled entirely by the mixed phase-space metric.


\subsection{Phase-sensitive Hall response of Dirac electrons}
As a concrete illustration of the general formalism, we consider a two-band model: $H = \boldsymbol{\sigma} \cdot \boldsymbol{h}(\boldsymbol{\xi}) + U(\boldsymbol{x})$, where $\boldsymbol{\sigma}$ denotes the vector of Pauli matrices, and $U(\boldsymbol{x})$ is a slowly varying scalar potential. Since $U(\boldsymbol{x})$ is proportional to the identity, it enters the local band energies $\Phieff_\pm = U(\boldsymbol{x}) \pm h(\boldsymbol{\xi})$ but does not modify the eigenstates. Hence, the quantum-geometric tensors are determined only by $\boldsymbol{h}(\boldsymbol{\xi})$. The Berry curvature of the lower band is \cite{Xiao10, Gao25}
\begin{equation}\label{eq:H2Omega}
    \Omega_{ij} (\boldsymbol{\xi}) = \frac{1}{2 h^3} \boldsymbol{h} \cdot (\partial_i \boldsymbol{h} \times \partial_j \boldsymbol{h}),
\end{equation}
and the band-renormalized quantum metric is \cite{Gao14, Gao25}
\begin{equation}\label{eq:H2G}
    G_{ij} (\boldsymbol{\xi}) = \frac{1}{4 h^3} [\partial_i \boldsymbol{h} \cdot \partial_j \boldsymbol{h} - (\boldsymbol{n} \cdot \partial_i \boldsymbol{h}) (\boldsymbol{n} \cdot \partial_j \boldsymbol{h})],
\end{equation}
where $h = \lvert \boldsymbol{h}\rvert$, and $\boldsymbol{n} = \boldsymbol{h}/h$ denotes the unit vector along $\boldsymbol{h}$. The metric-induced and $H_1$-induced corrections to the momentum-space Berry connection of the lower band are
\begin{eqnarray}
    A_i^{p\!\:(G)} &=& -G_{ij}^{pp}\partial_j^x U + G_{ij}^{pp}\partial_j^x h - G_{ij}^{px}\partial_j^p h, \label{eq:TwoBandAG}\\
    A_i^{p\!\:(H_1)} &=& -G_{ij}^{pp}\partial_j^x h. \label{eq:TwoBandAH1}
\end{eqnarray}
For this two-band model, the $H_1$-induced correction can itself be expressed in terms of the momentum-space metric $G_{ij}^{pp}$ and exactly cancels the $G_{ij}^{pp}\partial_j^x h$ term in $A_i^{p\!\:\smash{(G)}}\!$. Thus, the combined correction reduces to
\begin{equation}\label{eq:TwoBandAtot}
    A_i^{p\!\:(G)} + A_i^{p\!\:(H_1)} = -G_{ij}^{pp}\partial_j^x U - G_{ij}^{px}\partial_j^p h.
\end{equation}
The resulting momentum-space Berry curvature correction contributes to the intrinsic Hall response.


We consider a two-dimensional Dirac model in the presence of a magnetic texture and a scalar potential,
\begin{eqnarray}
    \boldsymbol{h} &=& v_\mathrm{F} (p_x,p_y,0) - J_\mathrm{ex}\left(0,m_y\cos(Kx),m_z\right)\!, \label{eq:Dirach}\\
    U &=& u \cos(Kx + \varphi). \label{eq:DiracU}
\end{eqnarray}
Here, $v_\mathrm{F}$ is the Fermi velocity, $J_\mathrm{ex}$ the exchange coupling, and $K$ the wave number. The oscillating in-plane component of the magnetic texture, $m_y\cos(Kx)$, shifts the local Dirac point to $p_y^*(x) = J_\mathrm{ex} m_y\cos(Kx)/v_\mathrm{F}$. The uniform out-of-plane magnetization $m_z$ acts as a Dirac mass and opens a gap $2|J_\mathrm{ex} m_z|$. The scalar potential has amplitude $u$ and phase shift $\varphi$ relative to the texture (see Fig.~\ref{fig:Fig2}(a)).

We work at zero temperature and choose the chemical potential such that $\mu< -|u| -|J_\mathrm{ex} m_z|$. This ensures that the system remains metallic throughout the modulation period. We further assume $|u|,|J_\mathrm{ex}m_z|\ll|\mu|$ and retain terms up to first order in $u$ and $m_z$. For the Dirac model in Eqs.~(\ref{eq:Dirach}) and (\ref{eq:DiracU}), the $G_{ij}^{pp}$ term in \eqref{eq:TwoBandAtot} gives no Hall contribution. This follows because the corresponding momentum-space integrand is odd under inversion about the local Dirac point. The combined $G$- and $H_1$-induced Hall conductivity is therefore entirely determined by the mixed phase-space metric:
\begin{equation}\label{eq:ModelG}
	\sigma_{xy}^{(G,\!\:\mathrm{mix})} = -\frac{e^2 J_\mathrm{ex} K v_\mathrm{F}}{16\pi \lvert \mu \rvert^3}m_y u\sin\varphi \!\;.
\end{equation}
This contribution vanishes when the magnetic texture and scalar potential are in phase or antiphase and reaches its maximum magnitude when the relative phase is $\varphi = \pm\pi/2$ (see Fig.~\ref{fig:Fig2}(b)). The phase dependence can be understood by interpreting the position-dependent shift $p_y^*(x)$ of the local Dirac point as an effective vector potential, $\mathcal{A}_y(x) = -p_y^*/e \propto - m_y\cos(Kx)$. The curl of this vector potential defines an emergent field, $\mathcal{B}_z(x) = \partial_x\mathcal{A}_y \propto K m_y\sin(Kx)$. This field should not be interpreted as driving the response through an ordinary Lorentz force: the mixed-metric Hall contribution remains intrinsic. The relevance of $\mathcal{B}_z$ instead follows from symmetry: like the Hall conductivity, $\mathcal{B}_z$ is a time-reversal-odd pseudoscalar in two dimensions. At the leading order, the local Hall contribution is linear in $\mathcal{B}_z$. Although $\mathcal{B}_z$ averages to zero over one period, the spatial average entering the physical response is weighted by the local equilibrium distribution. Under $x \to -x$, $\mathcal{A}_y$ and the energy contribution $h$ are even. For $\varphi=0$ or $\pi$, the scalar potential $U$ is even as well, so the equilibrium occupations at $x$ and $-x$ are equal. Since $\mathcal{B}_z(-x) = -\mathcal{B}_z(x)$, the two local contributions cancel. For any other relative phase, $U$ is not even under $x \to -x$, leading to unequal occupations and preventing complete cancellation.

It is useful to compare \eqref{eq:ModelG} with the anomalous Hall conductivity arising from the ordinary momentum-space Berry curvature. For the metallic regime considered here,
\begin{equation}\label{eq:ModelOmega}
	\sigma_{xy}^{(\Omega)} = \frac{e^2 J_\mathrm{ex}}{4\pi \hbar\lvert \mu \rvert} m_z.
\end{equation}
The conventional anomalous Hall conductivity is proportional to the uniform out-of-plane magnetization, whereas the mixed-metric Hall response depends on the in-plane magnetic texture, the scalar potential, and their relative phase. The latter can therefore remain finite as $m_z \to 0$, when the conventional anomalous Hall conductivity vanishes (see Fig.~\ref{fig:Fig2}(c)).


\subsection{Physical setting and estimated Hall response}
One possible physical setting is the surface of a three-dimensional strong topological insulator in contact with a magnetic layer. At the interface, the magnetic proximity effect induces an exchange coupling between the surface electrons and the spatially varying magnetization \cite{Nomura10, Luo13, Araki17}. The standard Hamiltonian for a single Dirac cone on this surface is equivalent to the form used here under a uniform rotation of the spin basis.
To estimate the magnitude of the mixed-metric Hall response, we take $v_\mathrm{F} = 5\times10^5\,\mathrm{m/s}$, typical of bismuth-chalcogenide topological insulators such as $\mathrm{Bi_2Se_3}$. First-principles calculations indicate that the magnetic proximity effect at a magnetic-insulator/$\mathrm{Bi_2Se_3}$ interface can induce an effective exchange field of tens of $\mathrm{meV}$ \cite{Luo13}. We adopt $J_\mathrm{ex} m_y = 20\,\mathrm{meV}$. Proximity-induced magnetic interactions have recently been investigated experimentally in $\mathrm{Cu_2OSeO_3/Bi_2Se_3}$ heterostructures \cite{Mehboodi26}. Adding a spatially varying component $m_x(x)$, as in a spiral texture, does not modify $\sigma_{xy}^{\smash{(G,\!\:\mathrm{mix})}}\!$ at the order considered. We therefore choose $\lambda = 2\pi/K=60\,\mathrm{nm}$, taking the bulk spiral pitch of the chiral magnetic insulator $\mathrm{Cu_2OSeO_3}$ as a representative magnetic length scale \cite{Adams12}. For the scalar potential, we choose a modulation amplitude $u = 20\,\mathrm{meV}$. Such a periodic potential could be generated electrostatically using a patterned dielectric \cite{Cano21}. Finally, we assume $\lvert\mu\rvert = 100\,\mathrm{meV}$, corresponding to a metallic surface state away from the Dirac point. At $\varphi = \pi/2$, these parameters yield a mixed-metric Hall conductivity of order $10^{-3}\!\:e^2/h$ (see Fig.~\ref{fig:Fig2}). Its characteristic dependence on the relative phase $\varphi$ distinguishes it from the conventional anomalous Hall contribution.

\section{Conclusions}
We have developed an effective semiclassical description of nonadiabatic wave-packet dynamics in full phase space, valid up to second order in $\hbar$. By encoding the nonadiabatic contributions as corrections to the Berry connection and wave-packet energy, this formulation enables the systematic derivation of the phase-space measure and kinetic equation. Retaining the full phase-space geometry reveals contributions to equilibrium and transport responses that cannot be captured by treating real- and momentum-space geometries separately. The quantum-metric correction to the measure yields a contribution to the electric polarization, while the effective momentum-space Berry curvature gives rise to an intrinsic linear Hall response. In the Dirac model, the Hall contributions from the momentum-space metric and the gradient correction cancel each other. The remaining Hall response is governed solely by the mixed phase-space metric and can remain finite even when the conventional anomalous Hall conductivity vanishes. The full phase-space metric thus emerges as an active ingredient in the equilibrium and transport properties of spatially inhomogeneous electronic systems.

\vspace{0.2em}
\paragraph{{\bf Acknowledgements} ---}
We acknowledge Y. Ren, S.-Y. Xu and K. Sun for fruitful discussions.
K.S. acknowledges financial support from the RIKEN Special Postdoctoral Researcher (SPDR) Program. 
N.N. was supported by JSPS KAKENHI Grant Numbers 24H00197, 24H02231 and 24K00583.
N.N. was supported by the RIKEN TRIP initiative.


%

\clearpage

\section{End Matter}
\subsection{Effective dynamics and phase-space measure}
Varying the Lagrangian in \eqref{eq:L} gives the equations of motion \cite{Yoshida25, Ren25A, Ren25B, Mameda25}
\begin{equation}\label{eq:Eq}
	\hbar^2 \left(G_{kj} \ddot{\xi}^j + \Gamma_{kij} \dot{\xi}^i \dot{\xi}^j\right) - \hbar \Omega_{kj} \dot{\xi}^j + J_{kj} \dot{\xi}^j + \partial_k \Phieff = 0,
\end{equation}
where $\Gamma_{kij} = (\partial_i G_{kj} + \partial_j G_{ki} - \partial_k G_{ij})/2$ is the Christoffel symbol and $\Omega_{kj} = \partial_{k}A_{j} - \partial_{j}A_{k}$ the Berry curvature. We seek a perturbative solution by expanding the phase-space velocity in powers of $\hbar$, $\dot{\xi}^{i} = \dot{\xi}_0^{i} + \hbar\, \dot{\xi}_1^{i} + \hbar^2 \dot{\xi}_2^{i}$. Substituting this expansion into \eqref{eq:Eq} and solving order by order in $\hbar$, we obtain
\begin{eqnarray}
	\dot{\xi}_0^{i} &=& -J^{ik} \partial_{k} \Phieff, \label{eq:Eq1}\\
	\dot{\xi}_1^{i} &=& -J^{ia} \Omega_{ab} J^{bk} \partial_{k} \Phieff, \label{eq:Eq2}\\
	\dot{\xi}_2^{i} &=& -J^{ia} \Omega_{ab} J^{bc} \Omega_{cd} J^{dk} \partial_{k} \Phieff - J^{ia} J^{jb} G_{ab} \partial_t \partial_j \Phieff \nonumber\\&&- J^{ia} J^{jb} J^{kc} \left(\Gamma_{abc}  \partial_{k} \Phieff + G_{ac} \partial_{b} \partial_{k} \Phieff\right)\partial_{j} \Phieff , \label{eq:Eq3}
\end{eqnarray}
where $J^{ik}$ is the inverse symplectic matrix: $J^{ik}J_{kj} = \delta^i_{\,j}$.

We now show that the effective Lagrangian in \eqref{eq:Leff} yields the same equations of motion up to $\mathcal{O}(\hbar^2)$. Up to $\mathcal{O}(\hbar)$, $L_{\mathrm{eff}}$ coincides with the original Lagrangian, \eqref{eq:L}, and therefore reproduces Eqs. (\ref{eq:Eq1}) and (\ref{eq:Eq2}). At order $\hbar^2$, the effective Lagrangian gives
\begin{eqnarray}\label{eq:Eq3eff}
	\dot{\xi}_{2,\mathrm{eff}}^{i} &=& -J^{ia} \Omega_{ab} J^{bc} \Omega_{cd} J^{dk} \partial_{k} \Phieff - J^{ik} \partial_t A_{k}^{(G)} \nonumber\\&& -J^{ia} \Omega_{ab}^{(G)} J^{bk} \partial_k \Phieff  - J^{ik} \partial_k \Phieff^{(G)},
\end{eqnarray}
where $\Omega_{ab}^{(G)} = \partial_{a}A_{b}^{(G)} - \partial_{b}A_{a}^{(G)}\!$ and we keep $\partial_t A_{k}^{(G)}\!$, as $A_{k}^{(G)}\!$ inherits the explicit time dependence of the wave-packet energy $\Phieff(t,\boldsymbol{\xi})$. Matching the terms containing an explicit time derivative in Eqs. (\ref{eq:Eq3}) and (\ref{eq:Eq3eff}) directly determines $A_{i}^{\smash{(G)}}\!$. Matching the remaining terms then determines $\Phieff^{\smash{(G)}}\!$, yielding the quantum-metric corrections in \eqref{eq:Corr}. Thus, up to $\mathcal{O}(\hbar^2)$, the effective Lagrangian in \eqref{eq:Leff} provides an equivalent description of the nonadiabatic wave-packet dynamics.

Once the nonadiabatic contributions are fully absorbed into $A_i^\mathrm{eff}\!$ and $\Phieff^\mathrm{eff}\!$, the phase-space density of states takes the standard form \cite{Xiao05},
\begin{equation}\label{eq:EqD}
	D = \frac{1}{(2\pi\hbar)^{d}} \!\left(\sqrt{\mathrm{det}(\hbar\!\:\Omega^\mathrm{eff} - J)} + \mathcal{O}\!\left(\hbar^{3}\right)\!\right)\!.
\end{equation}
Expanding the determinant up to second order in $\hbar$ yields \eqref{eq:EqDhbar2}.
The second-Chern-form contribution appearing there can be written as \cite{Zhao21}
\begin{equation}
     \mathcal{I}_{\Omega}^{(2)} = - \frac{1}{2} \!\left(\Omega_{ij}^{pp}\Omega_{ij}^{xx} + \Omega_{ij}^{px}\Omega_{ji}^{px} + \Omega_{ii}^{px}\Omega_{jj}^{xp}\right)\!,
\end{equation}
where the indices run over either real or momentum space, with superscripts specifying the corresponding space. The expression for $D$ in \eqref{eq:EqDhbar2} satisfies Liouville's theorem up to $\mathcal{O}(\hbar^2)$. Indeed, using the equations of motion, \eqref{eq:EqEff}, one can explicitly verify that $\dot{D} + D \partial_i\dot{\xi}^i = \mathcal{O}(\hbar^3)$.

\subsection{Pure momentum-space limit}
To benchmark our formalism against known results, we consider the case in which only momentum-space geometry is present. The equations of motion become
\begin{eqnarray}
	\dot{x}_i &=& v_i + e \hbar \Omega_{ik}^{pp} E_{k} + e^2 \hbar^2 \Gamma_{ijk}^{ppp} E_j E_k \nonumber\\
    & & -e \hbar^2 G_{ik}^{pp} \left(\partial_t + v_j \partial_{j}^{x}\right)\! E_k + \mathcal{O}\!\left(\hbar^3\right)\!,\label{eq:EqQ1}\\
    \dot{p}_i &=& -eE_i, \label{eq:EqQ2}
\end{eqnarray}
where $v_i = \partial_{i}^{p} \Phieff_0$ is the group velocity and, in this specific case, the unperturbed
band energy $\Phieff_0$ is independent of $\boldsymbol{x}$.
The density of states takes the form
\begin{equation}\label{eq:EqDQ}
	D = \frac{1}{(2\pi\hbar)^{d}} \left(1 +e \hbar^2 G_{jk}^{pp} \partial_{j}^{x} E_k  + \mathcal{O}\!\left(\hbar^{3}\right)\!\right)\!.
\end{equation}
In agreement with Ref.~\cite{Mameda25}, the quantum metric modifies the density of states through a term proportional to $\partial_{\boldsymbol{x}} \boldsymbol{E}$. Hence, it enters the current $j_i = -e\int{\!d^d p}\, D \dot{x}_i f$ with the following intrinsic DC contribution, up to $\mathcal{O}(\hbar^2)$,
\begingroup\setlength{\arraycolsep}{0pt}\begin{eqnarray}
	j_{i}^{(G)}\! = \!\int\!\frac{d^d p}{(2\pi\hbar)^{d}} & & \left[\frac{e^3 \hbar^2}{2} \!\left(2\partial_i^p G_{jk}^{pp} - \partial_j^p G_{ik}^{pp} - \partial_k^p G_{ij}^{pp}\right)\! E_j E_k \right. \nonumber\\
    & & \left. + e^2\hbar^2 \!\left(G_{ik}^{pp} v_j -  G_{jk}^{pp} v_i\right)\! \partial_{j}^{x} E_k \right]\! f_0, \qquad \label{eq:Eqj0}
\end{eqnarray}\endgroup
This result agrees with Ref.~\cite{Mameda25} and, remarkably, takes the same form as the current due to field-induced corrections \cite{Gao14}. This correspondence arises because, in this specific case, the nonadiabatic corrections in \eqref{eq:Corr} coincide with their field-induced counterparts. The first line in \eqref{eq:Eqj0} describes the nonlinear response \cite{Gao14, Gao15, Qiang25}, whereas the second line captures the contribution of the electric-field gradient \cite{Lapa19, Kozii21}, which has been associated with orbital magnetization \cite{Xiao21}.

We next examine the contribution of the term involving $\partial_t \boldsymbol{E}$. For a spatially uniform electric field that oscillates in time, $E_i(t) = \tilde{E}_i \,\mathrm{Re}(e^{-i\omega t})$, the current can be written as $j_i = \mathrm{Re}(j_{i}^{\smash{(0)}} + j_{i}^{\smash{(\omega)}} e^{-i\omega t} + j_{i}^{\smash{(2\omega)}} e^{-i2\omega t})$ with the coefficients
\begin{eqnarray}
	j_{i}^{(\omega)(\partial_t E)} \!&=&\! \tilde{E}_k  \!\int{\!\frac{d^d p}{(2\pi\hbar)^{d}} (-i\omega e^2 \hbar^2) G_{ik}^{pp} f_0}, \label{eq:EqjDt1}\\
	j_{i}^{(0)(\partial_t E)} \!&=&\! \tilde{E}_j \tilde{E}_k \int{\!\frac{d^d p}{(2\pi\hbar)^{d}} \frac{i\omega \tau e^3\hbar^2}{2(1-i\omega\tau)} G_{ik}^{pp} \partial_{j}^{p} f_0}, \qquad \label{eq:EqjDt2}
\end{eqnarray}
and $j_{i}^{(2\omega)(\partial_t E)} = -j_{i}^{(0)(\partial_t E)}$, where all contributions that do not descend from $\partial_t \boldsymbol{E}$ are omitted. Both the linear and the nonlinear responses include a term proportional to the driving frequency $\omega$. In linear response, the corresponding current oscillates with frequency $\omega$, whereas in nonlinear response it exhibits both a DC component and a second-harmonic component. The linear contribution in \eqref{eq:EqjDt1} is consistent with the Kubo formula and, in Ref.~\cite{Komissarov24}, has been associated with the intrinsic capacitance.

\subsection{Field-dependent measure and wave-packet energy}
We now return to the general phase-space geometry and allow the unperturbed band energy $\Phieff_0$ to depend on both $\boldsymbol{x}$ and $\boldsymbol{p}$. The electric field enters the modified phase-space density of states, \eqref{eq:EqDhbar2}, through
\begin{equation}\label{eq:EqDhbarMix}
	D^{(E)} = \frac{e\hbar^2 }{(2\pi\hbar)^{d}} \!\left(\partial_i^x G_{ij}^{pp} -\partial_i^p G_{ij}^{xp} + G_{ij}^{pp} \partial_i^x\right)\! E_j
\end{equation}
and the $G_{ij}$-induced energy correction, \eqref{eq:Corr}, through
\begin{equation}\label{eq:EqDMix}
	\Phieff^{(E)} = \frac{\hbar^2}{2} \!\left(e^2 G_{ij}^{pp} E_i +2e G_{ij}^{pp} \partial_i^x\Phieff_0 - 2 e G_{ij}^{xp} v_i\right)\! E_j.
\end{equation}
These corrections determine the contribution quadratic in the electric field in \eqref{eq:EqEE},  as well as the terms involving the quadrupole-moment density $Q_{ij}$, defined in \eqref{eq:Q}. In contrast, the polarization in \eqref{eq:PG} is determined by the field-independent part of $A_i^{p\!\:\smash{(G)}}\!$, obtained from \eqref{eq:Corr} by setting $\Phieff = \Phieff_0$. Similarly, the Hall response in \eqref{eq:sigmaG} is governed by the corresponding field-independent correction to the momentum-space Berry curvature.

We now give the explicit expression for the energy density.
Indices without superscripts, and not associated with the electric field, run over the full phase space:
\begin{widetext}
\begingroup\setlength{\arraycolsep}{0pt}\begin{eqnarray}
     \rho_{\varepsilon,\!\:\mathrm{tot}} = \!\int\!\frac{d^{d} p}{(2\pi\hbar)^{d}} & & \bigg\{\Phieff_0 f_0\!\left[1 + \hbar\!\:\Omega_{ii}^{px} + \hbar^2 \mathcal{I}_{\Omega}^{(2)}
    -\frac{\hbar^2}{2} \partial_i\!\left(J^{ia}G_{ab}J^{bk}\partial_k \Phieff_0\right)\right] - \frac{\hbar^2}{2} \partial_i\!\left(\Phieff_0 f_0 J^{ia}G_{ab}J^{bk}\partial_k \Phieff_0\right) \nonumber\\
    & & + e\hbar^2 \partial_i^x \!\left(\Phieff_0 f_0 G_{ij}^{pp}E_j\right) + \frac{\hbar^2}{2} \!\left(e^2 G_{ij}^{pp} E_i + e\!\:g_{ij}^{pp} \partial_i^x\right)\! E_j \!\:\bigg(1 + \Phieff_0 \frac{\partial}{\partial\Phieff_0}\bigg) f_0\bigg\} - e n\Phi. \label{eq:Etot}
\end{eqnarray}\endgroup
\end{widetext}

\subsection{Additional gradient-induced corrections}
For completeness, we report the corrections to the Berry connection and wave-packet energy generated by the interband matrix elements of the gradient perturbation $H_1$ within the formalism of Ref.~\cite{Ren25B}:
\begin{eqnarray}
    A_{i}^{(H_1)} &=& -\sum_{n\neq0} \frac{A_{i}^{0n}H_1^{n0} + \mathrm{c.c.}}{\Phieff_n - \Phieff_0}, \label{eq:Agr} \\ \Phieff^{(H_1)} &=& - \sum_{n\neq0} \frac{H_1^{0n}H_1^{n0}}{\Phieff_n - \Phieff_0}, \label{eq:Enegr}
\end{eqnarray}
where $0$ labels the band of interest and $n\neq0$ the remaining bands, $A_{i}^{nm}$ is the interband Berry connection, and
\begin{eqnarray}\label{eq:H1n0}
    H_{1}^{n0} &=& \frac{1}{2} \bigg[\!\left(V_{i}^{nn} -V_{i}^{00}\right) A_{i}^{p,n0} \nonumber\\ 
    & & + \sum_{l\neq0,n} \!\left(A_{i}^{p,nl} V_{i}^{l0} + V_{i}^{nl} A_{i}^{p,l0}\right)\!\bigg],
\end{eqnarray}
with $V_i^{nm}$ the $nm$ matrix element of $\partial_i^x \hat H$ evaluated at the wave-packet center. The correction to the Berry curvature is $\Omega_{ij}^{\smash{(H_1)}} = \partial_{i}A_{j}^{\smash{(H_1)}} - \partial_{j}A_{i}^{\smash{(H_1)}}\!$.

\end{document}